\documentclass[float,aps,prc,superscriptaddress,showpacs,onecolumn,notitlepage]{revtex4-1}

\usepackage{bm}       % bold math 
\usepackage{amsmath}
\usepackage{color} 
\usepackage{graphicx} % Include figure files 
\usepackage{epstopdf}
\usepackage{bbm}
\usepackage{multirow}
\usepackage{siunitx}

\newcommand {\nc} {\newcommand}

\nc {\IR} [1]{\textcolor{red}{#1}}
\nc {\IB} [1]{\textcolor{blue}{#1}}
\nc {\IP} [1]{\textcolor{magenta}{#1}}
\nc {\IM} [1]{\textcolor{Bittersweet}{#1}}
\nc {\IE} [1]{\textcolor{Plum}{#1}}

\nc{\ninej}[9]{\left\{\begin{array}{ccc} #1 & #2 & #3 \\ #4 & #5 & #6 \\ #7 & #8 & #9 \\ \end{array}\right\}}
\nc{\sixj}[6]{\left\{\begin{array}{ccc} #1 & #2 & #3 \\ #4 & #5 & #6 \\ \end{array}\right\}}
\nc{\threej}[6]{ \left( \begin{array}{ccc} #1 & #2 & #3 \\ #4 & #5 & #6 \\ \end{array} \right) }
\nc{\half}{\frac{1}{2}}
\nc{\numberthis}{\addtocounter{equation}{1}\tag{\theequation}}
\nc{\lla}{\left\langle}
\nc{\rra}{\right\rangle}
\nc{\lrme}{\left|\left|}
\nc{\rrme}{\right|\right|}

\begin{document}

\title{Nuclear spin features relevant to \textit{ab initio} nucleon-nucleus elastic scattering}

\author{R.~B.~Baker}
\affiliation{Institute of Nuclear and Particle Physics, and Department of Physics and Astronomy, Ohio University, Athens, OH 45701, USA}

\author{M.~Burrows}
\affiliation{Institute of Nuclear and Particle Physics, and Department of Physics and Astronomy, Ohio University, Athens, OH 45701, USA}
\affiliation{Department of Physics and Astronomy, Louisiana State University, Baton Rouge, LA 70803, USA}

\author{Ch.~Elster}
\affiliation{Institute of Nuclear and Particle Physics, and Department of Physics and Astronomy, Ohio University, Athens, OH 45701, USA}

\author{K.~D.~Launey}
\affiliation{Department of Physics and Astronomy, Louisiana State University, Baton Rouge, LA 70803, USA}

\author{P.~Maris}
\affiliation{Department of Physics and Astronomy, Iowa State University, Ames, IA 50011, USA}

\author{G.~Popa}
\affiliation{Institute of Nuclear and Particle Physics, and Department of Physics and Astronomy, Ohio University, Athens, OH 45701, USA}

\author{S.~P.~Weppner}
\affiliation{Natural Sciences, Eckerd College, St. Petersburg, FL 33711, USA}

\date{February 1, 2021}

\begin{abstract}
\begin{description}
\item[Background] Effective interactions for elastic nucleon-nucleus scattering from first principles require the use of the same nucleon-nucleon interaction in the structure and reaction calculations, as well as a consistent treatment of the relevant operators at each order.

\item[Purpose] Previous work using these interactions has shown good agreement with available data. Here, we study the physical relevance of one of these operators, which involves the spin of the struck nucleon, and examine the interpretation of this quantity in a nuclear structure context.

\item[Methods] Using the framework of the spectator expansion and the underlying framework of the no-core shell model, we calculate and examine spin-projected, one-body momentum distributions required for effective nucleon-nucleus interactions in $J=0$ nuclear states.

\item[Results] The calculated spin-projected, one-body momentum distributions for $^4$He, $^6$He, and $^8$He display characteristic behavior based on the occupation of protons and neutrons in single particle levels, with more nucleons of one type yielding momentum distributions with larger values. Additionally, we find this quantity is strongly correlated to the magnetic moment of the $2^+$ excited state in the ground state rotational band for each nucleus considered.

\item[Conclusions] We find that spin-projected, one-body momentum distributions can probe the spin content of a $J=0$ wave function. This feature may allow future \textit{ab initio} nucleon-nucleus scattering studies to inform spin properties of the underlying nucleon-nucleon interactions. The observed correlation to the magnetic moment of excited states illustrates a previously unknown connection between reaction observables such as the analyzing power and structure observables like the magnetic moment.

\end{description}
\end{abstract}

%\pacs{24.10.-i,24.10.Ht,25.40.-h,25.40.Cm}

\maketitle

\section{Introduction}
\label{sec:intro}

%%%%%%%%%%%%%%%%%%%%%%%%%%%%%%%%%%%%
% Introduction section
%%%%%%%%%%%%%%%%%%%%%%%%%%%%%%%%%%%%

The study of atomic nuclei is dependent on nuclear reactions to extract both reaction- and structure-related observables. From a theoretical perspective, one way to study these nuclear reactions is by reducing the many-body problem to a few-body problem and isolating the relevant degrees of freedom \cite{Johnson:2019sps}. This few-body problem can then be solved with the use of effective interactions, which are often called optical potentials. While different techniques have been implemented to construct these effective interactions from first principles, e.g.~Refs.~\cite{Rotureau:2016jpf,Burrows:2020qvu}, here we focus on the use of the \textit{ab initio} no-core shell model (NCSM) \cite{Navratil:2000ww,Roth:2007sv,BarrettNV13} and symmetry-adapted no-core shell model (SA-NCSM) \cite{LauneyDD16,Dytrych:2020vkl} to provide the relevant structure inputs. Specifically, we combine one-body densities for the target calculated from these methods with scattering approaches formulated to use structure and reaction input on an equal footing in a systematic expansion. For elastic scattering of protons and neutrons from nuclei, a microscopic optical potential can be derived with a Watson expansion of the multiple scattering series \cite{Crespo:1992zz,Crespo:1990zzb,Elster:1996xh,Elster:1989en,Arellano:1990xu,Arellano:1990zz}. This spectator expansion allows for the use of the same nucleon-nucleon ($NN$) interaction when calculating the one-body densities which must be folded with the $NN$ scattering amplitudes. By using realistic $NN$ and three-nucleon ($3N$) interactions derived from chiral effective field theory \cite{Machleidt:2011zz, Epelbaum:2014sza, Epelbaum:2014efa, Reinert:2017usi, Epelbaum:2019kcf}, we can implement this procedure in a fully \textit{ab initio} way, provided we include all relevant terms in the spectator expansion at each order.

Recent work \cite{Burrows:2020qvu} was able to construct and implement effective nucleon-nucleus interactions that include the spin of the struck target nucleon consistently at leading order. The results of using those effective interactions to study a nucleon elastically scattering off selected nuclei in $J=0$ states yielded only small changes in some of the spin observables when compared to previous work where the spin of the struck target nucleon was neglected. However, the pattern in where those changes occurred suggests that a closer investigation is required. Specifically, the ``closed shell" nuclei $^4$He and $^{16}$O exhibited no changes in their elastic scattering results and the $N=Z$ nucleus $^{12}$C showed only minimal changes. In contrast, the deviations in the analyzing power $A_y$ and spin-rotation parameter $Q$ for proton elastic scattering on $^6$He and $^8$He were much larger and it is a goal of this paper to examine the cause of these deviations in greater detail.

Progress in \textit{ab initio} nuclear structure, both in terms of the development of more specialized realistic interactions \cite{Ekstrom13, Shirokov:2016ead, Entem:2017gor, Jiang:2020the} and numerical improvements in the many-body methods \cite{doi:10.1002/cpe.3129, SHAO20181, LangrDDT18, LangrDDLT19}, have illustrated the large extent to which first principles calculations can describe nuclear states. In particular, the properties of excited states are increasingly well described from first principles, including collectivity in light to medium-mass nuclei \cite{LauneyDD16,Dytrych:2020vkl} and the emergence of rotational bands \cite{Dytrych:2007sv, DytrychSBDV_PRCa07,Maris:2014jha,Stroberg:2015ymf, Jansen:2015ngw,McCoy:2020xhp}. Indeed, \textit{ab initio} calculations have shown that only a few equilibrium shapes dominate within low-lying states, and that members of a rotational band have the same shape(s) and, in addition, exhibit very similar spin content \cite{DytrychSBDV_PRCa07,Dytrych:2020vkl}. This corroborates earlier studies, starting with the pioneering work of Refs.~\cite{Elliott58,RoweW2010book} and including large shell-model calculations \cite{DraayerWR84,BahriR00,PopaHD00}. That the nature of rotational bands can provide insights into orbital angular momentum and spin components of nuclear wave functions is also shown in Refs.~\cite{Johnson15, Caprio:2019yxh}. This is of particular importance here because the nuclear spin properties are often probed by a nuclear observable such as the magnetic moment, which is zero for $J=0$ states. In this paper, we exploit the similarity of the spin content within members of a rotational band, and by calculating magnetic moments for the first excited $2^+$ states, one can probe correlations between the spin features of the target detected by its magnetic moment and those detected by the spin-projection momentum distribution. Closely correlated results would imply that one can readily use the spin-projected momentum distribution in $J=0$ states to inform spin features in these states, or that measured magnetic moments can inform spin properties of effective interactions.

In this work, we seek to expand on the formalism presented in our previous work \cite{Burrows:2020qvu} and provide more physical insight into the effects of explicitly including the spin of the struck target nucleon in the effective interaction. In Sec.~\ref{sec:theory} we discuss the relevant derivations for the leading-order effective interaction, with a focus on the spin-dependent terms that arise. In Sec.~\ref{sec:results} we show results for these spin-dependent terms in the He isotope chain ($^4$He, $^6$He, and $^8$He) and discuss their physical interpretations. Furthermore, we note a correlation between these spin-dependent terms and a more traditional spin-related observable, the magnetic moment. We discuss our conclusions in Sec.~\ref{sec:conclusions}.

%%%%%%%%%%%%%%%%%%%%%%%%%%%%%%%%%%%%%%%%%%%%%%%%%%%%%%%%%%%%%%%%%

\section{Theoretical Framework}
\label{sec:theory}

Calculating elastic nucleon-nucleus scattering observables in an {\it ab initio} fashion
requires  
the interaction between the projectile and the target nucleus.
In a recent work~\cite{Burrows:2020qvu} this effective interaction was derived and calculated
in leading order of the spectator expansion of multiple scattering theory for elastic
scattering of protons (neutrons) off a $0^+$ ground state in selected nuclei. 
Here explicit care
is taken so that the $NN$ interaction is treated on the same footing in the structure 
as well as the reaction part of the calculation.

\begin{figure}
\centering
\includegraphics[width=0.45\textwidth]{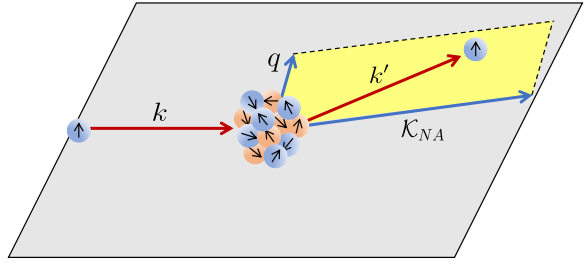}
\caption{Schematic diagram of the scattering plane, indicating the relevant momenta. Note that the normal vector $\hat{n}$ in Eq.~(\ref{eq2}) is perpendicular to both the $pp'$-plane and the $q\mathcal{K}$-plane.}
\label{fig1}
\end{figure}

For completeness we start with the explicit expression for the effective leading-order
interaction describing the scattering of a proton from a nucleus in a $0^+$ ground state, 
\begin{eqnarray}
\label{eq1}
\lefteqn{\widehat{U}_{\mathrm{p}}(\bm{q},\bm{\mathcal{K}}_{NA},\epsilon) =} & &  \cr
& & \sum_{\alpha=\mathrm{n,p}} \int d^3{\mathcal{K}} \eta\left( \bm{q}, \bm{\mathcal{K}}, \bm{\mathcal{K}}_{NA} \right) 
A_{\mathrm{p}\alpha}\left( \bm{q}, \frac{1}{2}\left( \frac{A+1}{A}\bm{\mathcal{K}}_{NA} - \bm{\mathcal{K}} \right); \epsilon
\right) \rho_\alpha^{K_s=0} \left(\bm{\mathcal{P}'}, \bm{\mathcal{P}}  \right) \cr  
&+&i (\bm{\sigma^{(0)}}\cdot\hat{\bm{n}}) \sum_{\alpha=\mathrm{n,p}} \int d^3{\mathcal{K}} \eta\left( \bm{q},
\bm{\mathcal{K}}, \bm{\mathcal{K}}_{NA} \right) 
C_{\mathrm{p}\alpha}\left( \bm{q}, \frac{1}{2}\left( \frac{A+1}{A}\bm{\mathcal{K}}_{NA} - \bm{\mathcal{K}} \right);
\epsilon
\right) \rho_\alpha^{K_s=0} \left(\bm{\mathcal{P}'}, \bm{\mathcal{P}}  \right) \cr 
&+&i \sum_{\alpha=\mathrm{n,p}} \int d^3{\mathcal{K}} \eta\left( \bm{q}, \bm{\mathcal{K}}, \bm{\mathcal{K}}_{NA}
\right) C_{\mathrm{p}\alpha} \left( \bm{q}, \frac{1}{2}\left( \frac{A+1}{A}\bm{\mathcal{K}}_{NA} - \bm{\mathcal{K}}
\right); \epsilon \right) S_{n,\alpha} \left(\bm{\mathcal{P}'}, \bm{\mathcal{P}} \right) \cos \beta\cr
&+&i (\bm{\sigma^{(0)}}\cdot\hat{\bm{n}}) \sum_{\alpha=\mathrm{n,p}} \int d^3{\mathcal{K}} \eta\left( \bm{q},
\bm{\mathcal{K}}, \bm{\mathcal{K}}_{NA} \right)  (-i) 
M_{\mathrm{p}\alpha} \left( \bm{q}, \frac{1}{2}\left( \frac{A+1}{A}\bm{\mathcal{K}}_{NA} - \bm{\mathcal{K}}
\right); \epsilon \right) S_{n,\alpha} \left(\bm{\mathcal{P}'}, \bm{\mathcal{P}}  \right) \cos \beta  .
\end{eqnarray}
The term $\eta\left( \bm{q}, \bm{\mathcal{K}}, \bm{\mathcal{K}_{NA}} \right)$  is the M{\o}ller
factor~\cite{CMoller} describing the transformation from the $NN$ frame to the $NA$ frame.
The functions $A_{\mathrm{p}\alpha}$, $C_{\mathrm{p}\alpha}$, and $M_{\mathrm{p}\alpha}$ represent the $NN$ interaction through Wolfenstein
amplitudes~\cite{wolfenstein-ashkin}. Since the incoming proton can interact
with either a proton or a neutron in the nucleus, the index $\alpha$ indicates the
neutron ($\mathrm{n}$) and proton ($\mathrm{p}$) contributions, which are calculated separately and then summed up. 
With respect to the nucleus, the operator $i (\bm{\sigma^{(0)}}\cdot \hat{\bm{n}})$ represents the spin-orbit operator in 
momentum space of
the projectile. As such, Eq.~(\ref{eq1}) exhibits the
expected form of an interaction between a spin-$\frac{1}{2}$ projectile and a target nucleus in a $J=0$ state \cite{RodbergThaler}. 
The momentum vectors in the problem are given as
\begin{eqnarray}
\label{eq2}
\bm{q} &=& \bm{p'} - \bm{p} = \bm{k'} - \bm{k}, \cr
\bm{\mathcal{K}} &=& \frac{1}{2} \left(\bm{p'} + \bm{p}\right), \cr
\hat{\bm{n}}&=&\frac{\bm{\mathcal{K}} \times \bm{q}}{\left| \bm{\mathcal{K}} \times
\bm{q}\right|} \cr
\bm{\mathcal{K}_{NA}} &=& \frac{A}{A+1}\left[\left(\bm{k'} + \bm{k}\right) +
      \frac{1}{2} \left(\bm{p'} + \bm{p}\right) \right], \cr
\bm{\mathcal{P}}&=& \bm{\mathcal{K}}+\frac{A-1}{A}\frac{\bm{q}}{2},  \cr
\bm{\mathcal{P'}}&=& \bm{\mathcal{K}}-\frac{A-1}{A}\frac{\bm{q}}{2}  .
\end{eqnarray}
A sketch of the scattering in the $NA$ frame is given in Fig.~\ref{fig1}, which includes the incoming momentum $\bm{k}$ of the projectile, its outgoing momentum $\bm{k'}$, the momentum transfer $\bm{q}$, and the average momentum $\bm{\mathcal{K}_{NA}}$. The struck nucleon in the target has an intial momentum $\bm{p}$ and a final momentum $\bm{p'}$. The two quantities representing the structure of the nucleus are the scalar one-body density
$\rho_\alpha^{K_s=0} \left(\bm{\mathcal{P}'}, \bm{\mathcal{P}}  \right)$ and the
spin-projected momentum distribution $S_{n,\alpha} \left(\bm{\mathcal{P}'}, \bm{\mathcal{P}}
\right)$. Both distributions are nonlocal and translationally invariant. 
Lastly, the term $\cos \beta$ in Eq.~(\ref{eq1}) comes from projecting $\bm{\hat{n}}$ from the $NN$ frame to the $NA$ frame. For further details, see Ref.~\cite{Burrows:2020qvu}.

The scalar one-body density is a well known quantity, while the spin-projected momentum
distributions have not been studied in detail. 
In general, a spin-dependent nonlocal density can be defined as~\cite{Burrows:2020qvu},
\begin{eqnarray}
\label{eq3}
\rho^{K_s}\left(\bm{p}, \bm{p}' \right) = \left\langle \Phi \left| \sum_{i=1}^{A} \delta^3( \bm{p_i} - \bm{p}) \delta^3( \bm{p_i}' - \bm{p}') \sum_{q_s} \sigma_{q_s}^{(i)K_s} \right| \Phi \right\rangle~,
\label{eqn5}
\end{eqnarray}
where $\sigma^{(i)K_s}_{q_s}$ is the spherical representation of the spin operator of the struck
nucleon in the target nucleus. 
The operator structure of the effective interaction~\cite{Golak:2010wz,Fachruddin:2000wv} given by Eq.~(\ref{eq1}) requires
we calculate the projections of the
spin operator on the momentum basis given by the vectors $\bm{q}$, $\bm{\mathcal{K}}$, and
$\bm{\hat{n}}$. 
Due to parity constraints the projections on  $\bm{q}$ and $\bm{\mathcal{K}}$ are zero. 
The projection along the $\hat n$-direction becomes
\begin{eqnarray}
\label{eq4}
S_n\left(\bm{p},\bm{p}' \right) \equiv \rho^{K_s} \left(\bm{p}, \bm{p}' \right) \cdot \hat{\bm{n}} = \sum_{q_s}{(-1)^{q_s}
\rho^{K_s=1}_{q_s} \left(\bm{p}, \bm{p}'\right)  \hat{\bm n}_{-q_s}},
\end{eqnarray}
where $\hat{n}$ has been written in terms of its spherical components. This equation can be explicitly evaluated as~\cite{Burrows:2020qvu}
\begin{eqnarray}
\label{eq5}
S_{n}(\bm{q},\bm{\mathcal{K}}) 
	&=& \sum_{q_s} (-1)^{-q_s} \sqrt{\frac{4\pi}{3}} Y^1_{-q_s}(\hat{\bm n}) \sum_{n_r l j n'_r l' j'} \sum_{K_l=|l-l'|}^{l+l'} \sum_{k_l=-K_l}^{K_l} \sum_{Kk} \lla K_l k_l, 1 q_s | K k \rra \lla J {-M}, K k | J {-M} \rra \cr
& & (-1)^{K} (-1)^{-l} \sqrt{\frac{3 (2j+1) (2j'+1) (2s+1) (2K_l+1)}{(2J+1)}} \ninej{l'}{l}{K_l}{s}{s}{1}{j'}{j}{K} \cr
& & \sum_{n_q,n_{\mathcal{K}},l_q,l_{\mathcal{K}}} 
\lla n_{\mathcal{K}} l_{\mathcal{K}}, n_q l_q : K_l | n'_r l', n_r l : K_l \rra_{d=1} R_{n_{\mathcal{K}} l_{\mathcal{K}}}(\mathcal{K}) R_{n_q l_q}(q) \mathcal{Y}_{K_l k_l}^{*l_q l_{\mathcal{K}}}(\widehat{\bm q},\hat{\bm{\mathcal{K}}}) \cr
& &\lla A \lambda J \left|\left| (a^{\dagger}_{n_r'l'j'} \tilde{a}_{n_rlj})^{(K)} \right|\right| A \lambda J \rra e^{\frac{1}{4A}b^2q^2}.
\end{eqnarray}
Note that $\lla n_{\mathcal{K}} l_{\mathcal{K}}, n_q l_q : K_l | n'_r l', n_r l : K_l \rra_{d=1}$ is a Talmi-Moshinsky bracket used to transform from the $pp'$ frame to the $q\mathcal{K}$ frame and the subscript $d=1$ is defined in Ref.~\cite{Burrows:2017wqn}. The factor $e^{\frac{1}{4A}b^2q^2}$ comes from removing the center-of-mass contributions in the one-body density matrix elements (OBDMEs) of the target. The OBDMEs, of which the $\left < A \lambda J | | (a^{\dagger}_{n_r'l'j'} \tilde{a}_{n_rlj})^{(K)} | | A \lambda J \right >$ term in Eq.~(\ref{eq5}) is the reduced form, are calculated as the inner product of the creation ($a^{\dagger}_{n_r l j m}$) and annhilation ($\tilde{a}_{n_r l j m} = (-1)^{j-m}a_{n_r l j {-m}}$) operators for single-particle harmonic oscillator states labeled by their $(n_r, l, j, m)$ values.

To facilitate calculations of the spin-projected momentum distribution, we make two choices: 1) use states with $J=0$ and 2) choose the vector $\hat{\bm q}$ in the z-direction and $\hat{\bm{\mathcal{K}}}$ in the x-z plane. This points $\hat{\bm{n}}$ along the negative y-direction, and Eq.~(\ref{eq5}) simplifies to
\begin{eqnarray}
\label{eq6}
	S_{n}(\bm{q},\bm{\mathcal{K}}) &=& (-i) \sum_{n_r l j n'_r l' j'} 
 \sqrt{2j+1} (-1)^{j+s+1} \sixj{l'}{l}{1}{s}{s}{j} \cr
& & \sum_{n_q,n_{\mathcal{K}},l_q,l_{\mathcal{K}}} \lla n_{\mathcal{K}} l_{\mathcal{K}}, n_q l_q : 1 | n'_r l', n_r l : 1 \rra_{d=1} R_{n_{\mathcal{K}} l_{\mathcal{K}}}(\mathcal{K}) R_{n_q l_q}(q) \sum_{q_s=-1,1} \mathcal{Y}_{1 -q_s}^{*l_q l_{\mathcal{K}}}(\widehat{\bm q},\hat{\bm{\mathcal{K}}}) \cr
& &\lla A \lambda 0 \left|\left| (a^{\dagger}_{n_r'l'j'} \tilde{a}_{n_rlj})^{(0)} \right|\right| A \lambda 0 \rra e^{\frac{1}{4A}b^2q^2}.
\end{eqnarray}
For $J=0$ states, the coupling coefficients require $K_s=1$ and $K_l=1$ in the Talmi-Moshinsky bracket. While we rely on Eq.~(\ref{eq6}) for the numerical implementation of $S_n(\bm{q},\bm{\mathcal{K}})$,
we can also examine one-dimensional functions depending only on the momentum transfer,
\begin{equation}
S_n(q) \equiv \int \mathrm{d}^3 {\mathcal{K}} S_n(\bm{q},\bm{\mathcal{K}}) =  \int \mathrm{d}\mathcal{K} \mathcal{K}^2 \int \mathrm{d}\theta_{q\mathcal{K}}
 \sin (\theta_{q\mathcal{K}}) \int \mathrm{d}\phi_{q\mathcal{K}} \: S_n(q, \mathcal{K}, \theta_{q\mathcal{K}}, \phi_{q\mathcal{K}}),
\label{eq7}
\end{equation}
and on the average momentum,
\begin{equation}
S_n(\mathcal{K}) \equiv \int \mathrm{d}^3 q S_n(\bm{q},\bm{\mathcal{K}}) =  \int \mathrm{d}q q^2 \int \mathrm{d}\theta_{q\mathcal{K}}
 \sin (\theta_{q\mathcal{K}}) \int \mathrm{d}\phi_{q\mathcal{K}} \: S_n(q, \mathcal{K}, \theta_{q\mathcal{K}}, \phi_{q\mathcal{K}}).
\label{eq8}
\end{equation}
Note that $\theta_{q\mathcal{K}}$ and $\phi_{q\mathcal{K}}$ are the polar and azimuthal angles between $\bm{q}$ and $\bm{\mathcal{K}}$. This definition is similar to defining the charge form factor, which is widely used to characterize momentum distributions in the nucleus. 
 
Lastly, it is worth noting the choice of specific coordinates in Eq.~(\ref{eq6}) prevents further analytical insights via integration, but we can use an alternate derivation of
Eq.~(\ref{eq4}) which yields~\cite{BurrowsM:2020}
\begin{eqnarray}
	S_n(\bm{q},\bm{\mathcal{K}}) 
	&=& -i \frac{3\sqrt{2}}{4\pi} \sum_{n_r l j n'_r l' j'} \sqrt{(2j+1)(2s+1)} (-1)^{j+s+1} \sixj{l'}{l}{1}{s}{s}{j} \cr
	& & \sum_{n_q,n_{\mathcal{K}},l_q,l_{\mathcal{K}}} \lla n_{\mathcal{K}} l_{\mathcal{K}}, n_q l_q : 1 | n'_r l', n_r l : 1 \rra_{d=1} R_{n_{\mathcal{K}} l_{\mathcal{K}}}(\mathcal{K}) R_{n_q l_q}(q) \cr
	& &\sum_w \hat{l}_q \hat{l}_{\mathcal{K}} \lla l_q 0 1 0 | w 0 \rra \lla l_{\mathcal{K}} 0 1 0 | w 0 \rra \sixj{l_q}{1}{w}{1}{l_{\mathcal{K}}}{1} (-)^{l_{\mathcal{K}}} \frac{P_w(\cos(\theta_{q\mathcal{K}}))}{|\sin(\theta_{q\mathcal{K}})|} \cr
	& & \lla A \lambda 0 \left|\left| (a^{\dagger}_{n_r'l'j'} \tilde{a}_{n_rlj})^{(0)} \right|\right| A \lambda 0 \rra e^{\frac{1}{4A}b^2q^2},
\label{eq9}
\end{eqnarray}
where $P_w(\cos(\theta_{q\mathcal{K}}))$ is a Legendre polynomial and the integer $w$ is determined by the angular momentum coupling. Furthermore, we can expand the sum over $w$ and consider the lowest nonzero term. This corresponds to $n_r=n'_r=0$ and $l=l'=1$, which only has one nonzero Talmi-Moshinsky bracket when $n_q=n_{\mathcal{K}}=0$ and $l_q=l_{\mathcal{K}}=1$. The Clebsch-Gordan coefficients in Eq.~(\ref{eq9}) are nonzero when $w=0,2$, yielding an angular dependence of the form
\begin{eqnarray}
S_n(\cos(\theta_{q\mathcal{K}})) \sim \frac{P_0(\cos(\theta_{q\mathcal{K}})) - P_2(\cos(\theta_{q\mathcal{K}}))}{|\sin(\theta_{q\mathcal{K}})|} = \frac{3}{2} \sin(\theta_{q\mathcal{K}}). 
\end{eqnarray}
Thus, while the full expression in Eq.~(\ref{eq9}) is complicated, in lowest order it has a relatively simple angular dependence.

%%%%%%%%%%%%%%%%%%%%%%%%%%%%%%%%%%%%%%%%%%%%%%%%%%%%%%%%%%%%%%%%%%%%%%%%%

\section{Results and Discussion}
\label{sec:results}

%%%%%%%%%%%%%%%%%%%%%%%%%%%%%%%%%%%%
% Results and Discussion section
%%%%%%%%%%%%%%%%%%%%%%%%%%%%%%%%%%%%

\begin{figure}
\begin{center}
\includegraphics[width=0.5\textwidth]{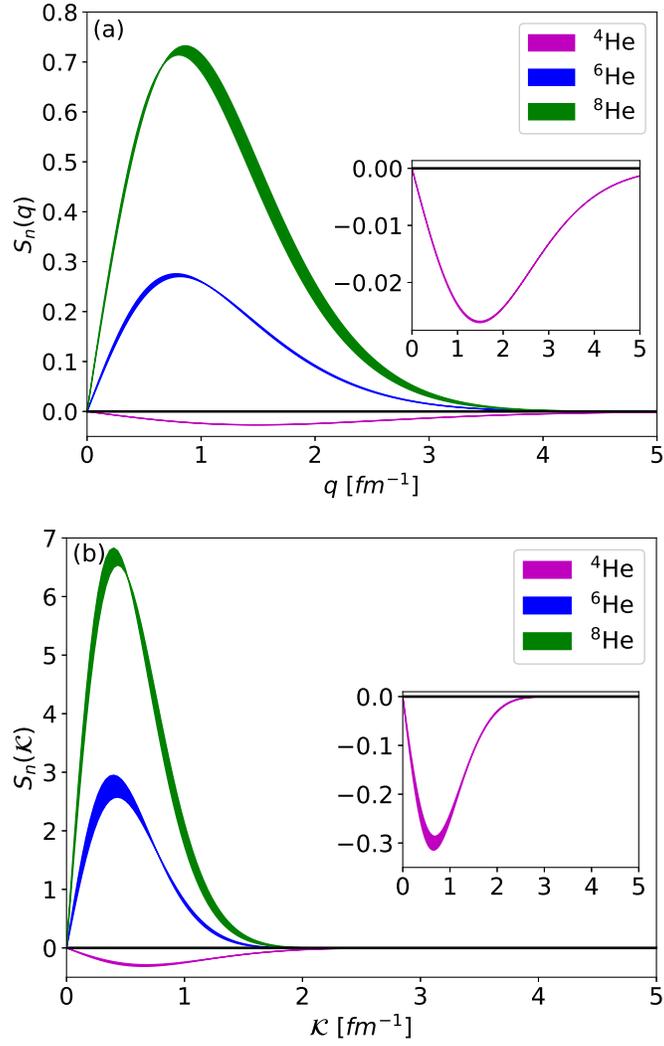}
\end{center}
\caption{(a) The function $S_n(q)$ for the neutron distribution in $^4$He at $N_{\mathrm{max}}=18$, $^6$He at $N_{\mathrm{max}}=18$, and $^8$He at $N_{\mathrm{max}}=14$ calculated with the NNLO$_{\mathrm{opt}}$ chiral interaction \cite{Ekstrom13}. (b) The function $S_n(\mathcal{K})$ for the same values. The bands in each plot indicate variations in $\hbar\Omega$ (16-24 MeV) at that value of $N_{\mathrm{max}}$. The insets show the $^4$He results in better detail.}
\label{fig2}
\end{figure}

First, we examine the functions $S_n(q)$ and $S_n(\mathcal{K})$, given by Eqs.~(\ref{eq7}) and (\ref{eq8}) respectively, for selected He isotopes. For these comparisons, we use the chiral interaction NNLO$_{\mathrm{opt}}$ \cite{Ekstrom13} and a large $N_{\mathrm{max}}$ value for each calculation, where $N_{\mathrm{max}}$ is the maximum number of harmonic oscillator quanta allowed above the valence shell for a given nucleus. For both $^6$He and $^8$He, their $S_n(q)$ and $S_n(\mathcal{K})$ behave similarly, though they differ in magnitude (Fig.~\ref{fig2}). In contrast, the curves for $^4$He are noticeably different, both in magnitude and sign. Notably, the $S_n(q)$ extend to a few fm$^{-1}$ while the $S_n(\mathcal{K})$ are largely concentrated below $1$ fm$^{-1}$. As $\mathcal{K}$ is considered a nonlocal variable, this likely reflects the nonlocality of $S_n(\bm{q},\bm{\mathcal{K}})$ is well confined to within the nucleus. While both can be informative, we focus on $S_n(q)$ due to its dependence on physically relevant momentum transfer, $q$. To better understand these differences and to develop expectations for what the $S_n(q)$ for any given nucleus might look like, we examine Eq.~(\ref{eq8}) more closely. As the only required inputs are the one-body density matrix elements (OBDMEs), we would expect $S_n(q)$ to have some dependence on the underlying shell structure for a given nucleus. 

\begin{figure}
\begin{center}
\includegraphics[width=0.5\textwidth]{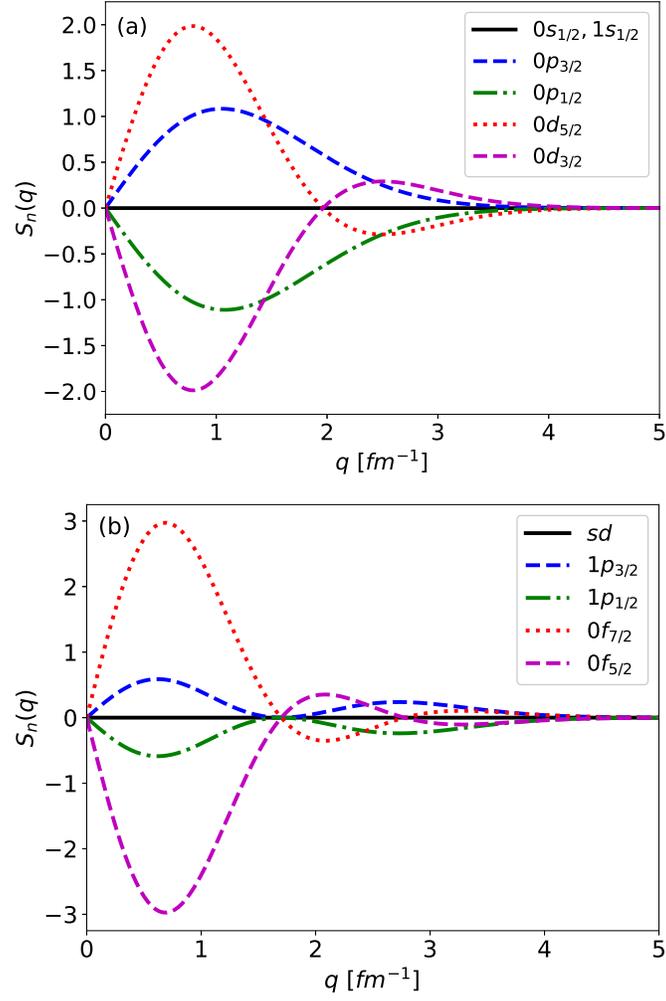}
\end{center}
\caption{The function $S_n(q)$ for protons or neutrons filling each closed shell configuration at $\hbar\Omega=20$ MeV for (a) the $s$, $p$, and $sd$ shell and (b) the $pf$ shell.}
\label{fig3}
\end{figure}

To study this in more detail, we performed toy model calculations in which each given harmonic oscillator state is completely filled and those nucleons are frozen. This results in there being zero probability of these nucleons moving to a different harmonic oscillator state, which means the resulting $S_n(q)$ is independent of the choice of nuclear interaction. This approach can then provide a basis for interpreting the $S_n(q)$ of a realisitic NCSM calculation. As such, in this toy model $S_n(q)$ displays characteristic behavior based on $n_r$, $l$, and $j$ values, as shown in Fig.~\ref{fig3}. For two nucleons frozen in the $0s_{1/2}$ shell, we find no contribution to $S_n(q)$ as expected for $l=l'=0$ in Eq.~(\ref{eq8}). For two nucleons frozen in the $0p_{1/2}$ shell, we find a contribution equal in magnitude but opposite in sign as four nucleons frozen in the $0p_{3/2}$ shell, despite there being more nucleons. In combination, this means full orbitals with the same $n_r,l$ values sum to zero $S_n(q)$ contributions, which can be seen for all HO shells up to $l=3$ in Fig.~\ref{fig3}. This pattern is always such that the $j=l+\frac{1}{2}$ state has an overall positive $S_n(q)$ and the $j=l-\frac{1}{2}$ state has an overall negative $S_n(q)$. Furthermore, the equal-in-magnitude behavior can be observed regardless of how many nucleons it takes to completely fill the shell, e.g. even for filled $0f_{5/2}$ (6 nucleons) versus $0f_{7/2}$ (8 nucleons), the magnitudes of $S_n(q)$ are still the same. Note that while the curves in Fig.~\ref{fig3} only show so called ``diagonal" OBDMEs, i.e. $a^{\dagger}_{\alpha} \tilde{a}_{\beta}$ where $\alpha=\beta$, these OBDMEs are larger than the off-diagonal OBDMEs where $\alpha \neq \beta$. Further, from Fig.~\ref{fig3} we can see that different $n_r$ values will yield different $S_n(q)$ curves, though they still maintain the same opposite-in-sign and equal-in-magnitude behavior.

\begin{figure}
\begin{center}
\includegraphics[width=\textwidth]{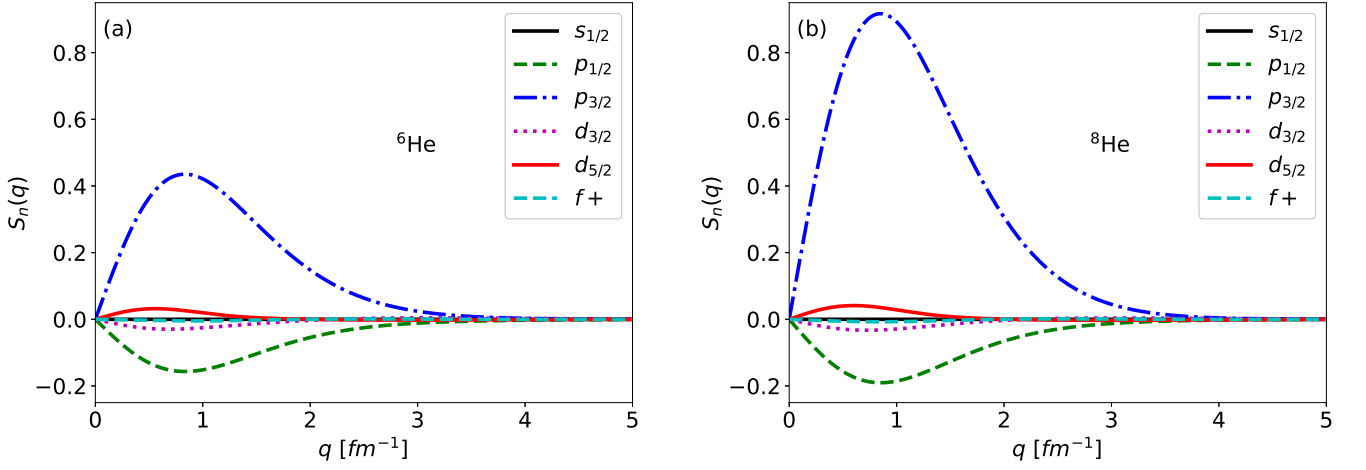}
\end{center}
\caption{The function $S_n(q)$ for the neutron distribution in (a) $^6$He at $N_{\mathrm{max}}=18$ and (b) $^8$He at $N_{\mathrm{max}}=14$, both with NNLO$_{\mathrm{opt}}$ \cite{Ekstrom13} and $\hbar\Omega=20$ MeV. The contributions to the total $S_n(q)$ curve are broken down by individual diagonal $lj$ OBDME values. See text for further discussion.}
\label{fig4}
\end{figure}

Given this information, we can now better interpret the $S_n(q)$ of a realistic NCSM calculation. Separating the OBDMEs to examine contributions to $S_n(q)$ for specific shells, we see a strong dominance of $p_{3/2}$ shells in $^6$He and $^8$He, as shown in Fig.~\ref{fig4}. Since we now know that filled shells have equal magnitude, we would expect similar partially-filled shells to have a similar magnitudes in their associated $S_n(q)$ curve. For both $^6$He and $^8$He we can see the $p_{3/2}$ curve has a magnitude of more than twice that of the $p_{1/2}$ curve, indicating nucleons prefer to fill the $p_{3/2}$ shells. This interpretation is supported by the ratio of the occupation probabilities calculated in the NCSM as well. Additionally, since we know the $n_r$ value changes the $S_n(q)$ curve, Fig.~\ref{fig4} also indicates the $0p$ shell largely dominates the total $S_n(q)$ since neither $^6$He nor $^8$He show the second peak indicative of the $1p$ shells.

Similarly, if we integrate $S_n(q)$ with respect to $q$,
\begin{eqnarray}
 S_n \equiv \int \mathrm{d}q q^2 S_n(q),
\label{eqn:Sn}
\end{eqnarray}
we can see that the $p_{3/2}$ contributions to the $S_n$ for the neutrons slightly more than doubles from $^6$He to $^8$He, as shown in Table \ref{tab1}. In contrast, the $p_{1/2}$ contributions to the $S_n$ for the neutrons barely changes. Note that Table \ref{tab1} also indicates that the proton contributions to $S_n$ barely change from $^6$He to $^8$He, suggesting that the proton information is mostly the same. The $S_n$ values for $^4$He are included in Table \ref{tab1} for comparison purposes. Specifically, it should be noted that the proton value for $^4$He is more negative than the proton values in $^6$He and $^8$He despite all three nuclei containing two protons. This possibly indicates the underlying OBDMEs include proton-neutron pairing effects for $N=Z$ nuclei largely suppressed when $N\neq Z$.

\begin{figure}
\centering
\includegraphics[width=0.5\textwidth]{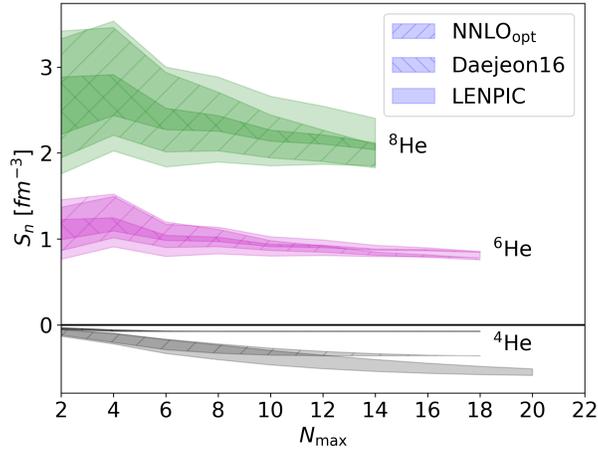}
\caption{Integrated $S_n$ values from Eq.~(\ref{eqn:Sn}) for neutrons in $^4$He, $^6$He, and $^8$He as calculated by the NNLO$_{\mathrm{opt}}$ \cite{Ekstrom13}, Daejeon16 \cite{Shirokov:2016ead}, and LENPIC-SCS at N2LO (NN potential) \cite{Epelbaum:2014sza, Binder:2015mbz, Binder:2018pgl} interactions. The band for each interaction corresponds to variations in $\hbar\Omega$.}
\label{fig5}
\end{figure}

While the function $S_n(q)$ clearly inherits information from the underlying shell structure, given its operator structure we would also expect it to provide more general information about the spin content of a given nucleus or nuclear interaction. To better facilitate those comparisons, we can examine the behavior of $S_n$ in more detail. For $^6$He and $^8$He, $S_n$ as a function of $N_{\mathrm{max}}$ is largely consistent for a variety of different nuclear interactions, as shown in Fig.~\ref{fig5} for NNLO$_{\mathrm{opt}}$ \cite{Ekstrom13}, Daejeon16 \cite{Shirokov:2016ead}, and LENPIC-SCS at N2LO (NN potential) \cite{Epelbaum:2014sza, Binder:2015mbz, Binder:2018pgl}. For high $N_{\mathrm{max}}$ values, we can start to see some slight deviations as the results approach convergence. For $^4$He, we can see clear differences in the value of $S_n$ depending on the choice of interaction. While Daejeon16 provides a result of almost zero, both NNLO$_{\mathrm{opt}}$ and LENPIC-SCS yield more negative values, both of which converge toward different values. This indicates the quantity $S_n$ probes a portion of nuclear interactions that remains distinct, even when the bulk observables, e.g.~binding energy and rms radius, would be in better agreement, e.g.~\cite{Shirokov:2016ead,Binder:2018pgl}.

While $S_n$ is not an observable -- it is not derived from a Hermitian operator -- we do find that it has a strong correlation with a well-defined observable: the magnetic moment. Specifically, when $S_n$ is calculated for the $0^+_{\mathrm{gs}}$ of a nucleus, we find a strong correlation between that value and the magnetic moment of the $2^+$ excited state in the ground state rotational band, e.g. $\mu = \lla 2^+ \left | M_1 \right | 2^+ \rra$. Using the impulse approximation to the magnetic moment, given as
\begin{eqnarray}
M_1 = \sqrt{\frac{3}{4\pi}} \mu_N \sum_{i=1}^A (g^{\ell}_i \ell_i + g^s_i s_i)
\label{eqn:mag_mom}
\end{eqnarray}
where $g^{\ell}_{\mathrm{p}} = 1$, $g^{\ell}_{\mathrm{n}} = 0$, $g^s_{\mathrm{p}} = 5.5857$, and $g^s_{\mathrm{n}} = -3.8263$ \cite{Suhonen}, we would expect some correlation since both operators explicitly include the spin operator. Additionally, given the properties of rotational bands, we would expect the spin components of those two wave functions to be quite similar since this $2^+$ state is a rotational ($L=2$) excitation of the $0^+_{\mathrm{gs}}$. To study these correlations, we employed a technique discussed in Ref.~\cite{LauneySDD_CPC14} and illustrated in Ref.~\cite{Burrows:2018ggt}. Briefly, we treat each calculation of $\mu$ and $S_n$ (for different values of $\hbar \Omega$ and $N_{\mathrm{max}}$) as elements of two separate vectors. The cosine of the angle between these two vectors, found by taking their inner product, tells us how much these quantities overlap and we can quantify this through a correlation coefficient $\zeta$. The sign of $\zeta$ refers to positive or negative correlation and the values span $|\zeta|=0$ (no correlation) to $|\zeta|=1$ (perfect correlation).

\begin{figure}
\begin{center}
\includegraphics[width=0.45\textwidth]{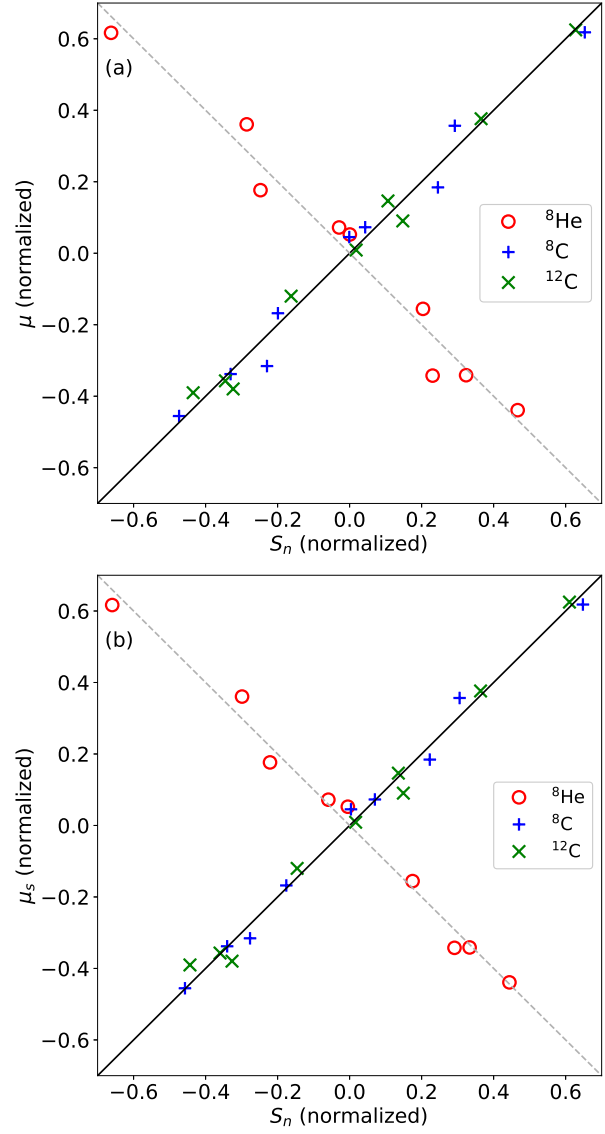}
\end{center}
\caption{(a) Correlation plot for the magnetic moment of the $2^+$ state in the ground state rotational band and the integrated total (protons + neutrons) $S_n$ value for the $0^+_{\mathrm{gs}}$ in $^8$He, $^8$C, and $^{12}$C calculated with NNLO$_{\mathrm{opt}}$. (b) Same plot for the spin contributions to the magnetic moment. Points for each nucleus were calculated at $\hbar\Omega=16,20,24$ MeV with $N_{\mathrm{max}}=6,8,10$ for $^8$He and $^8$C and $N_{\mathrm{max}}=4,6,8$ for $^{12}$C. The solid black line is a perfect positive correlation and the dashed gray line is a perfect negative correlation.}
\label{fig6}
\end{figure}

\begin{figure}
\begin{center}
\includegraphics[width=0.45\textwidth]{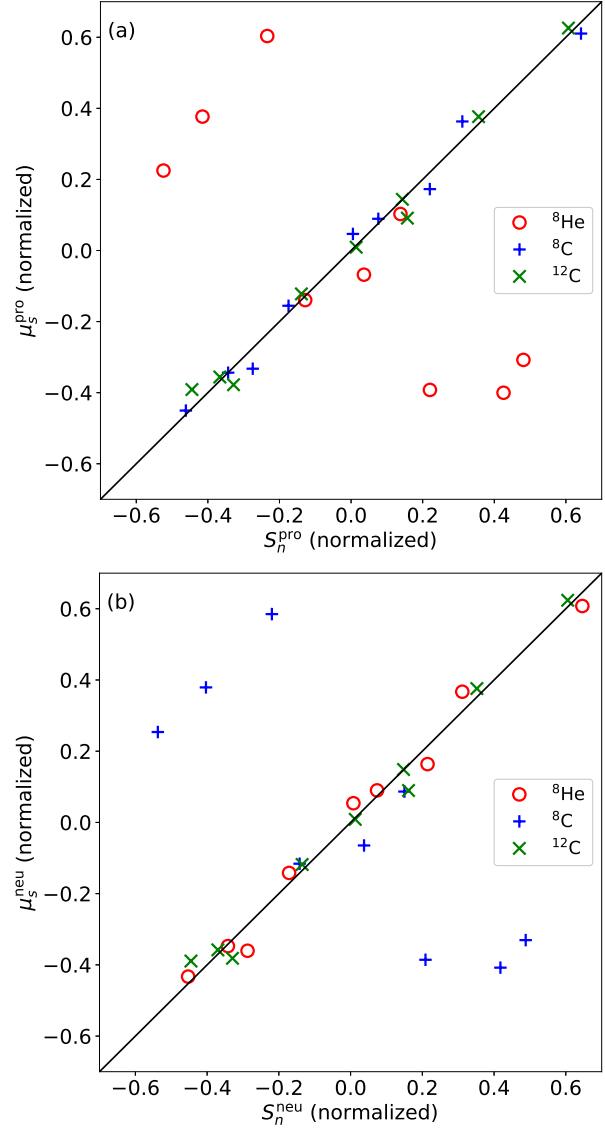}
\end{center}
\caption{(a) Correlation plot for the proton spin contributions to the magnetic moment of the $2^+$ state in the ground state rotational band and the integrated proton $S_n$ value for the $0^+_{\mathrm{gs}}$ in $^8$He, $^8$C, and $^{12}$C calculated with NNLO$_{\mathrm{opt}}$. (b) Same plot for the neutron spin contributions and the integrated neutron $S_n$ value. Points for each nucleus were calculated at $\hbar\Omega=16,20,24$ MeV with $N_{\mathrm{max}}=6,8,10$ for $^8$He and $^8$C and $N_{\mathrm{max}}=4,6,8$ for $^{12}$C. The solid black line is a perfect positive correlation.}
\label{fig7}
\end{figure}

Notably, in Fig.~\ref{fig6}(a), we see a strong positive correlation between the magnetic moment and $S_n$ for $N<Z$ ($\zeta^{^8\mathrm{C}}_{\mu,S_n}=0.990$) and $N=Z$ ($\zeta^{^{12}\mathrm{C}}_{\mu,S_n}=0.994$) nuclei, and a strong negative correlation for the $N>Z$ ($\zeta^{^8\mathrm{He}}_{\mu,S_n}=-0.983$) nucleus. This implies the spin information being probed by the magnetic moment of the $2^+$ excited state already exists in the wave function of the $0^+_{\mathrm{gs}}$ though it is not accessible to direct measurement. The strength of this correlation is the important factor, as the sign comes from the underlying $\mu$ and $S_n$ values, i.e.~both $\mu$ and $S_n$ are positive in $^8$C and $^{12}$C, but $\mu$ is negative and $S_n$ is positive in $^8$He. Additionally, if we separate the angular momentum contributions to the magnetic moment given by Eq.~(\ref{eqn:mag_mom}), we can see that these correlations are strongly driven by the spin terms in the magnetic moment, labeled as $\mu_s$ in Fig.~\ref{fig6}(b). When comparing Fig.~\ref{fig6}(a) and Fig.~\ref{fig6}(b), it can be seen that the orbital angular momentum terms in the magnetic moment slightly decrease the strength of these correlations, as the correlation coefficients for the latter plot are slightly larger ($\zeta^{^8\mathrm{C}}_{\mu_s, S_n}=0.996, \zeta^{^{12}\mathrm{C}}_{\mu_s, S_n}=0.995, \zeta^{^8\mathrm{He}}_{\mu_s, S_n}=-0.993$). Examining the spin components more closely, we can further separate it into proton and neutron contributions, as shown in Fig.~\ref{fig7}. The correlation between $\mu$ and $S_n$ is driven by the neutrons in $^8$He ($\zeta^{^8\mathrm{He}}_{\mu^{\mathrm{neu}}_s, S^{\mathrm{neu}}_n}=0.992$ versus $\zeta^{^8\mathrm{He}}_{\mu^{\mathrm{pro}}_s, S^{\mathrm{pro}}_n}=-0.791$), by the protons in $^8$C ($\zeta^{^8\mathrm{C}}_{\mu^{\mathrm{pro}}_s, S^{\mathrm{pro}}_n}=0.994$ versus $\zeta^{^8\mathrm{C}}_{\mu^{\mathrm{neu}}_s, S^{\mathrm{neu}}_n}=-0.803$), and shared equally by protons and neutrons in $^{12}$C ($\zeta^{^{12}\mathrm{C}}_{\mu^{\mathrm{pro}}_s, S^{\mathrm{pro}}_n}=0.995$ versus $\zeta^{^{12}\mathrm{C}}_{\mu^{\mathrm{neu}}_s, S^{\mathrm{neu}}_n}=0.994$).

%%%%%%%%%%%%%%%%%%%%%%%%%%%%%%%%%%%%%%%%%%%%%%%%%%%%%%%%%%%%%%%%%%%%%%%%%

\section{Conclusions and Outlook}
\label{sec:conclusions}

%%%%%%%%%%%%%%%%%%%%%%%%%%%%%%%%%%%%
% Conclusions section
%%%%%%%%%%%%%%%%%%%%%%%%%%%%%%%%%%%%

We have examined in detail the one-body, spin-projected momentum distribution $S_n(\bm{q}, \bm{\mathcal{K}})$, a new term that appears at leading order in the spectator expansion when the spin of the struck target nucleon is explicitly included. In particular, we have examined its dependence on the momentum transfer $q$ and showed changes in $S_n(q)$ for a given nucleus and for a given nucleon type in a nucleus. For the He-isotope chain ($^4$He, $^6$He, and $^8$He) the $S_n(q)$ for neutrons increase in magnitude as more neutrons are added, though the proton contributions in $^6$He and $^8$He remain the same and differ from those of $^4$He.

Noting the underlying shell structure inherent in our calculations, we identified interaction-independent characteristics of $S_n(q)$ based on which harmonic oscillator shells the nucleons occupy. This allowed for the interpretation of $S_n(q)$ for a realistic nucleus in better detail and showed that changes to $S_n(q)$ along the isotopic chain are related to which harmonic oscillator shells the subsequent nucleons are most probable to occupy. Furthermore, by investigating the dependence of the integrated quantity $S_n$ for different realistic nuclear interactions, we observed this behavior is largely independent of the interactions employed but noted different interactions may see yield slightly different overall values for $S_n$.

To better understand implications of this for an observable quantity, we compared calculations of $S_n$ for a $0^+_{\mathrm{gs}}$ to the magnetic moment of the $2^+$ state in that ground state rotational band and found a strong correlation, regardless of the nucleus. This correlation was driven by the spin components of the magnetic moment and indicates the quantity of interest here, $S_n$, (which is required to perform leading order calculations of effective interactions in a consistent way) probes spin information in a $J=0$ wave function that would normally be accessible only through observables in its excited states. This suggests the exciting possibility that future \textit{ab initio} nucleon-nucleus scattering studies could be sensitive to spin properties of the underlying nucleon-nucleon interaction, thereby providing further insight than previously appreciated.

%****************************************************************************

\begin{acknowledgments}
This work was performed in part under the auspices of the U.~S. Department of Energy under contract Nos. DE-FG02-93ER40756 and DE-SC0018223, and by the U.S. NSF (OIA-1738287 \& PHY-1913728). The numerical computations benefited from computing resources provided the Louisiana Optical Network Initiative and HPC resources provided by LSU, together with resources of the National Energy Research Scientific Computing Center, a DOE Office of Science User Facility supported by the Office of Science of the U.S. Department of Energy under contract No. DE-AC02-05CH11231.

\end{acknowledgments}

%%%%%%%%%%%%%%%%%%%%%%%%%%%%%%%%%%%%%%%%%%%%%%%%%%%%%%%%%%%%%%%%%%%%

\bibliographystyle{apsrev4-1}
\bibliography{denspot,clusterpot,ncsm}

\clearpage
%%%%%%%%%%%%%%%%%%%%%%%%%%%%%%%%%%%%%%%%%%%%%%%%%%%%%%
%---------- tables --------------

\begin{table}
\begin{center}
\begin{tabular}{| c | S[table-number-alignment = left] | S[table-number-alignment = left] | S[table-number-alignment = left] | S[table-number-alignment = left] | S[table-number-alignment = left] | S[table-number-alignment = left] |} \hline
\multirow{2}{*}{orbitals}	&	\multicolumn{6}{c|}{$S_n$ [fm$^{-3}$]}	\\											
	&	{$^4$He, protons}	&	{$^4$He, neutrons}	&	{$^6$He, protons}	&	{$^6$He, neutrons}	&	{$^8$He, protons}	&	{$^8$He, neutrons}	\\	\hline
$s_{1/2}$	&	0.000	&	0.000	&	0.000	&	0.000	&	0.000	&	0.000	\\	
$p_{1/2}$	&	-0.606	&	-0.607	&	-0.274	&	-0.537	&	-0.236	&	-0.606	\\	
$p_{3/2}$	&	0.150	&	0.149	&	0.154	&	1.271	&	0.124	&	2.565	\\	
$d_{3/2}$	&	0.169	&	0.169	&	0.074	&	0.070	&	0.040	&	0.049	\\	
$d_{5/2}$	&	-0.033	&	-0.032	&	-0.032	&	-0.036	&	-0.016	&	-0.034	\\	
$f+$	&	-0.043	&	-0.043	&	-0.018	&	-0.011	&	-0.016	&	-0.011	\\	\hline 
total	&	-0.362	&	-0.363	&	-0.096	&	0.757	&	-0.104	&	1.963	\\	\hline
\end{tabular}
\end{center}
\caption{Values of $S_n$ broken down by nucleon type and diagonal $lj$ OBDME values for NNLO$_{\mathrm{opt}}$ \cite{Ekstrom13} at $\hbar\Omega=20$ MeV. Results for $^4$He are at $N_{\mathrm{max}}=18$, for $^6$He at $N_{\mathrm{max}}=18$, and for $^8$He at $N_{\mathrm{max}}=14$. The neutron values for $^6$He and $^8$He calculated correspond applying Eq.~(\ref{eqn:Sn}) to each $S_n(q)$ in Fig.~\ref{fig4}. Note the row labeled $f+$ corresponds to all remaining diagonal OBDMEs and all off-diagonal ODBMEs. See text for further detail.}
\label{tab1}
\end{table}

\clearpage
%%%%%%%%%%%%%%%%%%%%%%%%%%%%%%%%%%%%%%%%%%%%%%%%%%%%%%
%-------  figures  ---------

\end{document}